\documentclass{appolb}
\usepackage{graphicx}

\begin{document}
\title{Quantum Mechanics and CPT tests with neutral kaons \\at the KLOE experiment%
\thanks{Presented at Symposium on applied nuclear physics and innovative technologies, Krak{\'o}w, 3-6.06.2013}%
}
\author{Izabela Balwierz-Pytko \\ on behalf of the KLOE-2 collaboration \footnote{\scriptsize{D.~Babusci, D.~Badoni, I.~Balwierz-Pytko, G.~Bencivenni, C.~Bini, C.~Bloise, F.~Bossi, P.~Branchini, A.~Budano, L.~Caldeira~Balkest\aa hl, G.~Capon, F.~Ceradini, P.~Ciambrone, F.~Curciarello, E.~Czerwi\'nski, E.~Dan\`e, V.~De~Leo, E.~De~Lucia, G.~De~Robertis, A.~De~Santis, P.~De~Simone, A.~Di~Domenico, C.~Di~Donato, R.~Di~Salvo, D.~Domenici, O.~Erriquez, G.~Fanizzi, A.~Fantini, G.~Felici, S.~Fiore, P.~Franzini, A.~Gajos, P.~Gauzzi, G.~Giardina, S.~Giovannella, E.~Graziani, F.~Happacher, L.~Heijkenskj\"old, B.~H\"oistad, L.~Iafolla, M.~Jacewicz, T.~Johansson, K.~Kacprzak, A.~Kupsc, J.~Lee-Franzini, B.~Leverington, F.~Loddo, S.~Loffredo, G.~Mandaglio, M.~Martemianov, M.~Martini, M.~Mascolo, R.~Messi, S.~Miscetti, G.~Morello, D.~Moricciani, P.~Moskal, F.~Nguyen, A.~Palladino, A.~Passeri, V.~Patera, I.~Prado~Longhi, A.~Ranieri, C.~F.~Redmer, P.~Santangelo, I.~Sarra, M.~Schioppa, B.~Sciascia, M.~Silarski, C.~Taccini, L.~Tortora, G.~Venanzoni, W.~Wi\'slicki, M.~Wolke, J.~Zdebik}}
\address{Institute of Physics, Jagiellonian University, Krak{\'o}w, Poland}
\\}
\maketitle
\begin{abstract}
Neutral kaons produced in the correlated pairs at the DAFNE $\phi$-factory offer unique possibilities to perform fundamental tests of CPT invariance, as well as of the basic principles of quantum mechanics. The analysis of the data collected by the KLOE experiment allows to improve results on several parameters describing CPT violation and decoherence and to measure the regeneration cross section on the beam pipe materials.
\end{abstract}
  
\section{The KLOE experiment at the DAFNE collider}

The DAFNE $\phi$-factory, located at the Frascati National Laboratory (LNF) of INFN, is an $e^+ e^-$ collider, working at the energy of the $\phi$ resonance $\sqrt{s}=m_{\phi}\approx1019$~MeV. The KLOE detector, placed at the center of one of the two interaction points, completed its first data taking compaign in March 2006. Collected data corresponds to $\sim2.5~\textrm{fb}^{-1}$ of integrated luminosity, which translates to $\sim2.5$ billion of $\phi$ meson decays into neutral kaon pairs.

The KLOE detector covers almost full solid angle and consists of a cylindrical drift chamber \cite{DC} with excellent momentum and vertex reconstruction, surrounded by an electromagnetic calorimeter \cite{EmC} with very good time resolution, both inserted in a superconducting coil which produces an axial magnetic field of 0.52~T, parallel to the beam axis. 

\section{Neutral kaon interferometry}

At KLOE neutral kaons are produced in the decay of the $\phi$ meson in a fully antisymmetric entangled state:
{\small
\begin{eqnarray}
\left|i\right>=\frac{N}{\sqrt{2}}\left[\left|K_S(+\vec{p})\right>\left|K_L(-\vec{p})\right>-\left|K_L(+\vec{p})\right>\left|K_S(-\vec{p})\right>\right],
\label{eq:initial_2}
\end{eqnarray}} 
\noindent where $N$ is a normalization factor. 

Due to the large lifetime difference of both kaons ($\tau_L\approx51$~ns, $\tau_S\approx90$~ps) there is also a large difference in their mean decay lengths, namely for $K_S$ it is about 6~mm whereas for $K_L$ about 3.5~m for energy range covered by KLOE. This fact enables to identify the $K_L$ meson decay by the presence of $K_S$ decay close to the interaction region.

The observable quantity is the double differential decay rate of the state in eq. (\ref{eq:initial_2}) into decay products $f_1$ and $f_2$ at the proper times $t_1$ and $t_2$, respectively. The decay intensity distribution as a function of the decay time difference $\Delta t=t_1-t_2$ between both kaon decays reads \cite{neutral_kaon}:
{\small
\begin{eqnarray}
I(f_1,f_2;\Delta t)= \frac{C_{12}}{\Gamma_S+\Gamma_L}\Big[|\eta_1|^2 e^{-\Gamma_L \Delta t}+|\eta_2|^2 e^{-\Gamma_S \Delta t} \nonumber \\
-2|\eta_1| |\eta_2| e^{-\frac{(\Gamma_S+\Gamma_L)}{2}\Delta t} \cos(\Delta m \Delta t+\Delta\varphi)\Big]
\label{eq:delta_t}
\end{eqnarray}}
with a phase difference $\Delta\varphi=\varphi_2-\varphi_1$ and:
{\small
\begin{eqnarray}
C_{12}=\frac{|N|^2}{2}\left|\left<f_1|T|K_S\right>\left<f_2|T|K_S\right>\right|^2, \ \ \eta_i=|\eta_i|e^{i\varphi_i}\equiv\frac{\left<f_i|T|K_L\right>}{\left<f_i|T|K_S\right>}.
\label{eq:eta}
\end{eqnarray}}
Eq. (\ref{eq:delta_t}) holds for $\Delta t \ge 0$, while for $\Delta t<0$ the substitutions $\Delta t\to|\Delta t|$ and $1\leftrightarrow 2$ have to be applied. Here, apart from the exponential decay terms of $K_L$ and $K_S$ we have also an interference term that is characteristic at $\phi$-factories. From this distribution for various final states $f_i$ one can determine directly: $\Gamma_S$, $\Gamma_L$, $\Delta m$, $\eta_i$, $\Delta\varphi$ and perform tests of CP and CPT symmetries comparing experimental distributions with the theoretical predictions.

\section{Search for decoherence and CPT violation in entangled neutral kaons}

If both $K_L$ and $K_S$ decay into any identical final states $f_1=f_2$, for example $K_L\to \pi^+ \pi^-$ and $K_S\to \pi^+ \pi^-$, from eq. (\ref{eq:eta}) can be seen that $\eta_1=\eta_2=\eta$ and $\varphi_1=\varphi_2$. Substituting this to eq. (\ref{eq:delta_t}) one obtains:
{\small
\begin{eqnarray}
I(f_1=f_2;|\Delta t|)= \frac{C_{12}|\eta|^2}{\Gamma_S+\Gamma_L}\Big[e^{-\Gamma_L |\Delta t|}+e^{-\Gamma_S |\Delta t|} -2e^{-\frac{(\Gamma_S+\Gamma_L)}{2}|\Delta t|} \cos(\Delta m |\Delta t|)\Big].
\label{eq:delta_t_2}
\end{eqnarray}}
The above equation implies that two kaons cannot decay into the same final states \emph{at the same time}. This counterintuitive correlation is of the type first pointed out by Einstein, Podolsky and Rosen in their famous paper \cite{EPR}.

 In general decoherence denotes the transition of a pure state into an incoherent mixture of states, meaning that entanglement of particles is lost. The decoherence parameter $\zeta$ can be introduced by multiplying the interference term in eq. (\ref{eq:delta_t_2}) by a factor $(1-\zeta)$ \cite{neutral_kaon}:
{\small
\begin{eqnarray}
I(\pi^+\pi^-,\pi^+\pi^-;\Delta t)\propto e^{-\Gamma_L \Delta t}+e^{-\Gamma_S \Delta t}- 2(1-\zeta_{SL})e^{-\frac{(\Gamma_S+\Gamma_L)}{2}\Delta t} \cos(\Delta m \Delta t).
\label{eq:decoherence}
\end{eqnarray}}
A value of $\zeta=0$ corresponds to the usual quantum mechanics case, $\zeta=1$ to the total decoherence and different values to intermediate situations between these two.

At KLOE tests of the coherence were performed by analyzing data corresponding to the $\sim$1.5~fb$^{-1}$ of integrated luminosity. The fit of eq. (\ref{eq:decoherence}) to the experimental distribution of the $\phi\to K_L K_S\to \pi^+\pi^-\pi^+\pi^-$ intensity as a function of the absolute value of $\Delta t$ was performed \cite{first_obs}. The results presented in ref. \cite{CPT_and_GM_tests} show no deviations from the quantum mechanics predictions:
{\small
\begin{eqnarray}
\zeta_{SL}=\left(0.3\pm1.8_{\mathrm{stat}}\pm 0.6_{\mathrm{syst}}\right)\cdot 10^{-2}, \ \ \zeta_{0\bar{0}}=\left(1.4\pm 9.5_{\mathrm{stat}}\pm 3.8_{\mathrm{syst}}\right)\cdot 10^{-7}.
\end{eqnarray}}
This result can be compared to the results obtained by collaborations CPLEAR \cite{PR}: $\zeta_{0\bar{0}}=0.4\pm0.7$ and BELLE (measured in the B meson system) \cite{PRL}: $\zeta_{0\bar{0}}^B=0.029\pm0.057$.

There are several hypothesis on a possible origin of the CPT violation. One of them is related to a possible decoherence in quantum gravity that induces an ill definition of the CPT operator. As indicated in ref. \cite{Bernabeu}, in this case the definition of the particle-antiparticle states has to be modified and the antisymmetric two-kaon state may get a small admixture of the symmetic state:
{\small
\begin{eqnarray}
\left|i\right>\propto(K_S K_L-K_L K_S)+\omega(K_S K_S-K_L K_L).
\end{eqnarray}}
The parameter $\omega$ is a complex CPT violation parameter that could be measured only in entangled systems. One expects that it is at most \cite{neutral_kaon}: $|\omega|^2=O\left(\frac{E^2 / M_{Planck}}{\Delta\Gamma}\right)\approx 10^{-5} \Rightarrow |\omega|\sim 10^{-3}$.

This analysis was performed by the KLOE collaboration on the same $I(\pi^+\pi^-\pi^+\pi^-;\Delta t)$ distribution as before by fitting the decay intensity distribution modified including the $\omega$ parameter. The obtained result is consistent with no CPT violation effects \cite{CPT_and_GM_tests}:
{\small
\begin{eqnarray}
\Re \omega=\left(-1.6^{+3.0}_{-2.1 \ \mathrm{stat}}\pm 0.4_{\mathrm{syst}}\right)\cdot 10^{-4}, \ \ \Im \omega=\left(-1.7^{+3.3}_{-3.0 \ \mathrm{stat}}\pm 1.2_{\mathrm{syst}}\right)\cdot 10^{-4}. 
\end{eqnarray}}
The upper limit at 95\% confidence level for the module is $|\omega|\le1.0 \cdot 10^{-3}$. In comparision, in the B meson system only the real part of it was estimated and with a limited precision \cite{Nebot}: $-0.0084\le\Re \omega\le 0.0100$.

\section{$K_S$ regeneration at KLOE}
At KLOE-2 \cite{EPJC} the statistical error on decoherence and CPTV parameters can be reduced because of higher luminosity and a new detector close to the interaction point: Cylindrical-GEM Inner Tracker \cite{Archilli:2010xb}. The main source of systematic errors is due to the poor knowledge of the incoherent regeneration process $K_L \to K_S\to\pi^+ \pi^-$ in the cylindrical beam pipe made of beryllium and located 4.3~cm from the interaction region. The corresponding systematic uncertainty can be largely reduced by improving the measurement of the incoherent regeneration cross section on the beam pipe materials \cite{thesis}. This process constitutes a background for decoherence and CPT violation searches in $K_L K_S\to \pi^+\pi^- \pi^+ \pi^-$ decays.\\

\textbf{Acknowledgments}\\
This work was supported in part by the European Commission under the 7th Framework Programme, Grant Agreement No. 283286; by the Polish National Science Centre through the Grants No. 0469/B/H03/2009/37, 0309/B/H03/2011/40, DEC-2011/03/N/ST2/02641, 2011/01/D/ST2/00748 and by the Foundation for Polish Science through the MPD programme and the project HOMING PLUS BIS/2011-4/3.


\end{document}